\documentclass[final,5p,times,twocolumn]{elsarticle}

\usepackage{amssymb}

\journal{Physica C}

\begin{document}

\begin{frontmatter}



\title{Quantum critical point for stripe order: \\ An organizing principle of cuprate superconductivity}


\author{Nicolas Doiron-Leyraud
and Louis Taillefer}

\address{D\'epartement de physique \& RQMP, Universit\'e de Sherbrooke, Sherbrooke, Canada \\
Canadian Institute for Advanced Research}

\begin{abstract}

A spin density-wave quantum critical point (QCP) is the central organizing principle of organic, iron-pnictide, 
heavy-fermion and electron-doped cuprate superconductors. 
It accounts for the superconducting $T_c$ dome, the non-Fermi-liquid resistivity, and the Fermi-surface reconstruction. 
Outside the magnetically ordered phase above the QCP, scattering and pairing decrease in parallel as the system moves away from the QCP. 
Here we argue that a similar scenario, based on a stripe-order QCP, is a central organizing principle of hole-doped cuprate superconductors. 
Key properties of La$_{1.6-x}$Eu$_{0.4}$Sr$_x$CuO$_4$, La$_{1.6-x}$Nd$_{0.4}$Sr$_x$CuO$_4$ and YBa$_2$Cu$_3$O$_y$ 
are naturally unified, including stripe order itself, its QCP, Fermi-surface reconstruction, the linear-$T$ resistivity,
and the nematic character of the pseudogap phase. 

\end{abstract}


\end{frontmatter}


\section{Introduction}
\label{Intro}

Quantum oscillations~\cite{NDLNature2007,BanguraPRL2008,YellandPRL2008,JaudetPRL2008,SebastianNature2008,AudouardPRL2009} 
and Hall effect~\cite{LeBoeufNature2007,LeBoeufPRB2011} experiments on underdoped YBa$_2$Cu$_3$O$_y$ (YBCO) have revealed a 
small electron Fermi surface, indicative of Fermi surface reconstruction (FSR) 
by density-wave order~\cite{ChakravartyScience2008,TailleferJPCM2009,VojtaPhysics2011}. 
The nature of the density-wave order, and its relation to the pseudogap and high-$T_c$ superconductivity, 
are now key issues for our understanding of cuprates~\cite{NormanPhysics2010}. 
We have recently examined this question by looking at the response caused by the FSR in the thermoelectric properties of a number of cuprates, 
establishing parallels and analogies between seemingly different materials. 
In this article, we first discuss materials where the presence of static stripe order is convincingly established by a large body of data. These are the family of doped LSCO materials, in particular La$_{1.6-x}$Eu$_{0.4}$Sr$_x$CuO$_4$ (Eu-LSCO) and La$_{1.6-x}$Nd$_{0.4}$Sr$_x$CuO$_4$ (Nd-LSCO), which exhibit static charge-stripe order. We then turn to YBCO and show that the thermoelectric response as a function of temperature and doping is essentially identical to that of Eu-LSCO, evidence that the same mechanism of FSR, namely stripe order, is at play in both materials. 
Finally, we examine the full ordering process upon cooling, starting with its onset at the pseudogap temperature $T^{\star}$. This is where, in YBCO, the Nernst effect shows the onset of a large in-plane anisotropy, revealing the nematic character of the pseudogap phase.

\section{Stripe order and FSR in Eu-LSCO}
\label{Stripe}

\begin{figure}[h!]
\center		 		
\includegraphics[scale=2.2]{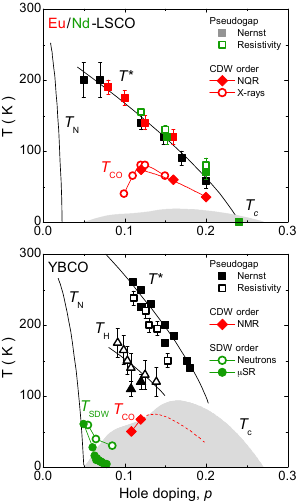}
\caption{Top: Temperature-doping phase diagram of Nd-LSCO and Eu-LSCO. 
$T_{\mathrm{CO}}$ denotes the onset of charge density-wave order in Eu-LSCO as measured by X-rays (red circles,~\cite{FinkPRB2011}) 
and NQR (red diamonds,~\cite{HuntPRB2001}). 
The pseudogap temperature $T^{\star}$ is defined in two ways: 
1) the temperature $T_{\rho}$ (open squares) at which the electrical resistivity deviates from its high-temperature linear behaviour 
and 2) the temperature $T_{\nu}$ (full squares) at which the Nernst coefficient expressed as $\nu/T$ deviates from its high temperature, weakly linear, regime. 
$T_{\rho}$ was extracted from resistivity data on Nd-LSCO~\cite{DaouNatPhys2009,IchikawaPRL2000}. 
$T_{\nu}$ was extracted from Nernst data on LSCO (black squares,~\cite{WangPRB2006}), Eu-LSCO (red squares,~\cite{OCCNature2009}), 
and Nd-LSCO (green squares,~\cite{OCCNature2009}), as reported in~\cite{TailleferAR2010,OCCpreprint2012}. 
Bottom: Temperature-doping phase diagram of YBCO. 
$T_{\mathrm{CO}}$ (red diamonds) is from NMR data~\cite{MarcHenriNature2011}. 
$T_{\rho}$ (open squares) is from an analysis~\cite{DaouNature2010} of data in Ref.~\cite{AndoPRL2004}. 
$T_{\nu}$ (full squares) is from Nernst data~\cite{DaouNature2010}. 
$T_{\rm H}$ is defined in the main text and is extracted in Ref.~\cite{LeBoeufPRB2011} from our own (full triangles) and previously published data 
(open triangles,~\cite{SegawaPRB2004}). 
$T_{\rm SDW}$ is the onset temperature for spin density-wave order as measured by neutron scattering~\cite{HaugNJP2010} and $\mu$SR~\cite{ConeriPRB2010}. 
In both panels, $T_{\rm N}$ is a schematic of the N\'eel temperature and $T_c$ is the superconducting transition temperature described by the grey dome, 
from data in Ref.~\cite{LiangPRB2006} for YBCO.
}
\label{Fig1} 
\end{figure}

In Figure~\ref{Fig1}, we show the temperature-doping phase diagram of Eu-LSCO. 
In this material, static charge stripe order has been observed by X-rays~\cite{FinkPRB2011} and NQR~\cite{HuntPRB2001} 
at a doping-dependent temperature $T_{\rm CO}$ that peaks at $p=1/8$, but remains sizable up to at least $p=0.20$. 
We emphasize that $T_{\rm CO}$ is well separated from other transitions, such as the structural transition that occurs near 130 K, 
or the superconducting $T_c$ which never exceeds 20 K. 
In Figure~\ref{Fig2}, we reproduce the X-ray intensity on Eu-LSCO at $p = 0.11$ and 0.125 as a function of temperature, 
which shows the onset of charge order at $T_{\rm CO}$. 
We also display our measurements of the Seebeck coefficient $S$ on Eu-LSCO~\cite{ChangPRL2010,FLNatureComm2011} at the very same doping values, 
expressed as $S$ over the temperature $T$, much like the linear-$T$ component of specific heat. 
We see that $S/T$ is small and positive at high temperature, and then drops to negative values. 
At low temperature, $S/T$ becomes large and negative, indicating that it is dominated by a small electron-like Fermi surface~\cite{BehniaJPCM2004}.
Taking the maximum in $S/T$ as a loose criterion for the temperature at which FSR occurs, the close correspondence between 
this temperature and $T_{\rm CO}$ is evidence that the FSR is caused by stripe order.
Furthermore, a negative $S/T$ at low temperature is only observed when stripe order is present, both being absent 
below $p = 0.08$, as we recently reported~\cite{FLNatureComm2011}.


\section{FSR and stripe order  in YBCO}

\begin{figure}[t]	
\center		 		
\includegraphics[scale=1.3]{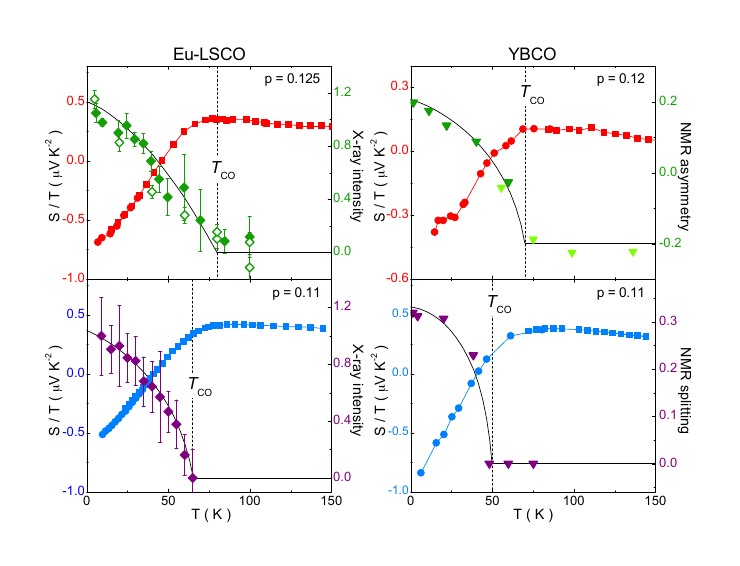}
\caption{Left: Seebeck coefficient $S$ expressed as $S/T$ as a function of temperature for Eu-LSCO at $p$=0.125 and 0.11, 
in zero field (squares) and 10 T (circles) (from~\cite{ChangPRL2010,FLNatureComm2011}). 
The X-ray scattering intensity as a function of temperature in Eu-LSCO is also shown for the corresponding dopings
(closed diamonds from~\cite{FinkPRB2011,FinkPRB2009} and courtesy
of J. Fink; open diamonds from~\cite{OCCNature2009}), 
revealing the onset of charge order at $T_{\mathrm{CO}}$ (vertical dashed lines). 
Right: $S/T$ as a function of temperature for YBCO at $p=0.12$ and 0.11, in zero field (squares) and 28 T (circles) (from~\cite{ChangPRL2010,FLNatureComm2011}). 
The NMR signal from charge order in YBCO at the corresponding dopings is also reproduced (dark green triangles for 30~T data, light green triangles for 33.5~T data, purple triangles for 28.5~T data, all from~\cite{MarcHenriNature2011}).
}
\label{Fig2} 
\end{figure}

As shown in Figure~\ref{Fig2}, $S/T$ in YBCO exhibits the very same drop to negative values at low temperatures~\cite{ChangPRL2010,FLNatureComm2011}. 
The magnitude of the negative residual value of $S/T$ at $T \to 0$ is in good agreement with the magnitude expected from the small Fermi surface pocket detected 
by quantum oscillations~\cite{NDLNature2007,JaudetPRL2008}, given that $S/T \propto m^{\star} / F$~\cite{BehniaJPCM2004}, 
where $F$ and $m^{\star}$ are the frequency and cyclotron mass of the oscillations~\cite{ChangPRL2010,FLNatureComm2011}, respectively.
The dependence of $S/T$ on both temperature and doping in Eu-LSCO and YBCO shows a detailed and striking similarity,
which lead us to conclude that the FSR in YBCO is also driven by stripe order~\cite{ChangPRL2010,FLNatureComm2011}. 
Recent nuclear magnetic resonance (NMR) experiments on YBCO confirmed this finding~\cite{MarcHenriNature2011}, revealing charge stripe order at a temperature $T_{\rm CO}$ 
which closely matches that at which the drop in $S/T$ occurs (see Figure~\ref{Fig2}). 
In Figure~\ref{Fig1}, we reproduce the onset temperature $T_{\rm CO}$ measured by NMR on the phase diagram of YBCO.

Note that stripe order in Eu-LSCO exists in the absence of a magnetic field, whereas for YBCO at the dopings mentioned here ($p=0.11$ and 0.12) 
static stripe order only appears when a sufficiently large magnetic field is applied. 
We attribute this difference to phase competition. 
In Eu-LSCO, owing perhaps to the more favourable crystal structure, stripe order is naturally stronger and suppresses superconductivity to a lower $T_c$. 
In YBCO, stripe order is weaker and superconductivity stronger, with a much larger $T_c$, therefore requiring a large magnetic field to suppress the latter and tip the balance in favour of stripe order which is otherwise suppressed.


\section{Quantum critical point in Nd-LSCO}
\label{QCP}

In Eu-LSCO, stripe order extends over a large portion of the temperature-doping phase diagram~\cite{FinkPRB2011,HuntPRB2001},
as seen from the doping dependence of the onset temperature $T_{\rm CO}$ in Figure~\ref{Fig1}. 
The very same phase diagram is observed for the closely related material Nd-LSCO~\cite{HuntPRB2001,NiemollerEurPhysJB1999}. 
In Nd-LSCO, we have tracked the transport properties up to the doping where stripe order vanishes, 
at the quantum critical point $p^{\star}$ = 0.24~\cite{OCCPhysicaC2010}. 
At this doping, we observed a strictly linear temperature dependence of the electrical resistivity (both in the plane and along the $c$-axis)~\cite{DaouNatPhys2009}, 
as well as a logarithmic divergence of the thermopower, $S/T\propto$ log($1/T$)~\cite{DaouPRB2009}, two archetypal signatures of a QCP~\cite{PaulPRB2001}.

This type of QCP has four principal consequences: 
1) Fermi-surface reconstruction; 
2) anomalous (non-Fermi-liquid) scattering;
3) unconventional pairing; and 
4) phase competition. 
All four consequences are clearly observed in organic~\cite{BourbonnaisPRB2009,NDLPRB2009}, pnictide~\cite{CanfieldAR2010}, 
heavy-fermion~\cite{MathurNature1998,MonthouxNature2007,KnebelCRAS2011} and electron-doped cuprate ~\cite{DaganPRL2004,ArmitageRMP2010} superconductors, 
where the order in all cases is spin-density-wave (SDW) order.
(In pnictides, the SDW order is stripe-like, {\it i.e.} unidirectional.)
The dome-like region of superconductivity in the phase diagram of these four families of materials -- the rise and fall of $T_c$ -- is due to 
pairing above, and competition below, the QCP, respectively. 
The linear-$T$ resistivity at the QCP, and transport anomalies below it, are due to scattering by spin fluctuations
and Fermi-surface reconstruction by SDW order, respectively. 
Moreover, the strength of the anomalous scattering is directly correlated with the pairing strength, in that the slope of the linear-$T$ resistivity scales with the superconducting $T_c$, both decreasing in parallel as one moves away from the QCP~\cite{TailleferAR2010}. This was observed in the organic Bechgaard salt (TMTSF)$_2$PF$_6$~\cite{NDLPRB2009}, 
the iron-pnictide Ba(Fe$_{1-x}$Co$_x$)$_2$As$_2$ (Co-Ba122)~\cite{NDLPRB2009}, and the electron-doped cuprate La$_{2-x}$Ce$_x$CuO$_4$ (LCCO)~\cite{JinNature2011}.

\begin{figure}[t]	
\center		 		
\includegraphics[scale=2.2]{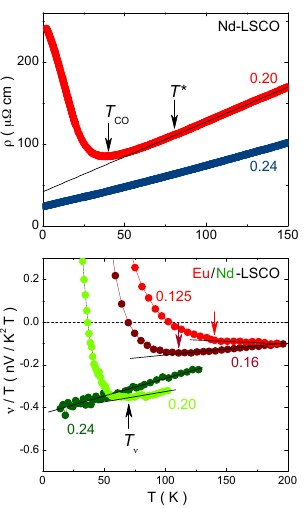}
\caption{Top: Electrical resistivity as a function of temperature for Nd-LSCO at $p=0.20$ and 0.24, in the normal state revealed by applying a 
magnetic field of 35 T~\cite{DaouNatPhys2009}. 
The line is a linear fit to the high-temperature data at $p=0.20$. 
The temperature at which the data deviate from this fit is $T_{\rho} \equiv T^{\star} = 80$~K. 
The onset of charge order detected by NQR at $T_{\rm CO} = 40$~K~\cite{HuntPRB2001} coincides with the minimum in the resistivity.
At $p=0.24$, the resistivity remains strictly linear in temperature as $T \to 0$. 
Bottom: Nernst coefficient expressed as $\nu/T$ as a function of temperature for Eu-LSCO (red circles) and Nd-LSCO (green circles), 
at doping values as indicated. 
The lines are linear fits to the data at high temperature and the arrows mark the temperature $T_{\nu}$ below which the data 
deviate upward from the fit~\cite{OCCNature2009}. 
In Nd-LSCO at $p=0.24$, no anomaly is seen in either the electrical resistivity or the Nernst effect.
}
\label{Fig3} 
\end{figure}

The QCP at which SDW order sets in is the central organizing principle with which to understand the overall phenomenology
of these superconductors.
We now propose that a QCP at which stripe order sets in is a central organizing principle of hole-doped cuprates.
With its QCP at $p^{\star}$ = 0.24, this principle readily applies to the case of Nd-LSCO (and Eu-LSCO).
 Fermi-surface reconstruction is observed in several transport properties as the doping is reduced below 
 $p^{\star}$~\cite{DaouNatPhys2009,OCCNature2009,DaouPRB2009}.
In Figure~\ref{Fig3}, the normal-state electrical resistivity of Nd-LSCO at $p$ = 0.20 is seen to exhibit
a pronounced upturn below 40~K~\cite{DaouNatPhys2009}, the temperature at which NQR detects the onset of 
stripe order~\cite{HuntPRB2001}.
By contrast, no anomaly is seen at $p = 0.24$.
What is seen instead is a perfectly linear-$T$ resistivity as $T \to 0$, when superconductivity is suppressed by a large magnetic field~\cite{DaouNatPhys2009}.
This is the signature of the anomalous scattering that occurs at the QCP.
The same linear-$T$ resistivity is observed at the SDW QCP of 
the electron-doped Pr$_{2-x}$Ce$_x$CuO$_4$ (PCCO)~\cite{FournierPRL1998} and LCCO~\cite{JinNature2011}, 
the organic metal (TMTSF)$_2$PF$_6$~\cite{NDLPRB2009},
and the pnictides Co-Ba122~\cite{FangPRB2009} and BaFe$_{2}$(As$_{1-x}$P$_{x}$)$_{2}$ (P-Ba122)~\cite{KasaharaPRB2010}.

\begin{figure}[h!]	
\center		 		
\includegraphics[scale=1.7]{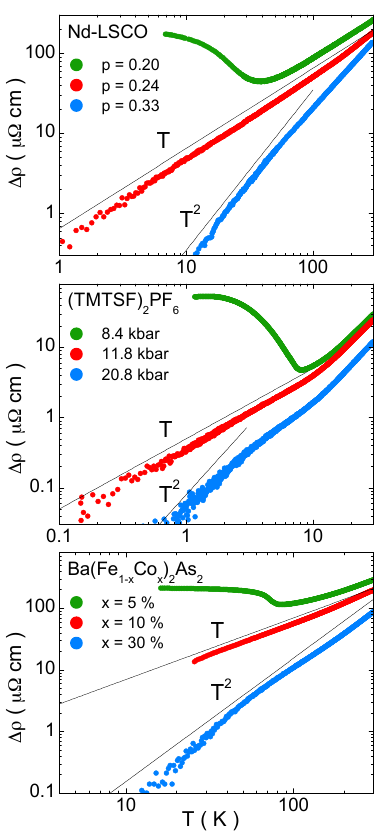}
\caption{Temperature-dependent part of the in-plane normal-state resistivity of materials in three families of superconductors, plotted as $\Delta\rho=\rho(T)-\rho_0$, where $\rho_0$ is the residual resistivity, versus temperature on a log-log scale. Three values of the relevant tuning parameter were chosen: below, at, and above their respective quantum critical points. Top: Data on hole-doped cuprates Nd-LSCO at $p=0.20$ and 0.24~\cite{DaouNatPhys2009} and LSCO $p=0.33$~\cite{NakamaePRB2003}. The QCP at a hole doping $p=0.24$ marks the end of the stripe-ordered phase in Nd-LSCO~\cite{DaouNatPhys2009,IchikawaPRL2000}. Figure adapted from Ref.~\cite{DaouPRB2009}. Middle: Data on the organic Bechgaard salt (TMTSF)$_2$PF$_6$. The QCP at a pressure $P=10$ kbar marks the end of the SDW phase. Figure adapted from Ref.~\cite{NDLPRB2009}. Bottom: Data on the iron-pnictide Ba(Fe$_{1-x}$Co$_x$)$_2$As$_2$~\cite{FangPRB2009}. The QCP at a nominal Co concentration $x=10 \%$ marks the end of the SDW phase.
}
\label{FigB} 
\end{figure}

In Figure~\ref{FigB}, we display the evolution of the electrical resistivity across three regimes typical of a quantum critical point, for a hole-doped cuprate, a Bechgaard salt, and an iron pnictide: a FSR below, linear-$T$ at, and Fermi-liquid behavior well above, the QCP. The similarity between materials coming from different families is striking and supports the universal character of a QCP scenario.

Two questions arise.
First, is this organizing principle of a stripe QCP universal amongst hole-doped cuprates?
While further work is required to answer that question, two universal features would suggest so:
1) all hole-doped cuprates have a dome of superconductivity, with an optimal $T_c$ at $p \simeq 0.16$;
2) near optimal doping, all hole-doped cuprates have a resistivity which is linear in temperature, with a universal slope when expressed per CuO$_2$ plane~\cite{NDLarxiv2009}. 
Moreover, as in the other families of superconductors discussed so far, that slope scales with the superconducting $T_c$ as a function of doping, with the linear term vanishing where superconductivity ends~\cite{TailleferAR2010,NDLarxiv2009,ManakoPRB1992,CooperScience2009}.
The second question is: How does the pseudogap phase fits into such a QCP scenario?
In the remainder, we explore that question.


\section{Pseudogap phase and nematicity}
\label{Nematic}

In Figure~\ref{Fig3}, we see that the upturn in the resistivity of Nd-LSCO at $p=0.20$ is a gradual one. There is no sharp anomaly at $T_{\rm CO} = 40$~K.
Cooling from high temperature, $\rho(T)$ deviates from its linear-$T$ dependence below $T_{\rho} \simeq 80$~K.
In other words, the resistivity senses the coming onset of stripe order well before long-range order actually sets in --
there is a wide precursor regime before Fermi-surface reconstruction is detected in the Hall or Seebeck coefficients, for example,
with $T_{\rho} \simeq 2~T_{\rm CO}$.

\begin{figure}[t]	
\center		 		
\includegraphics[scale=2.2]{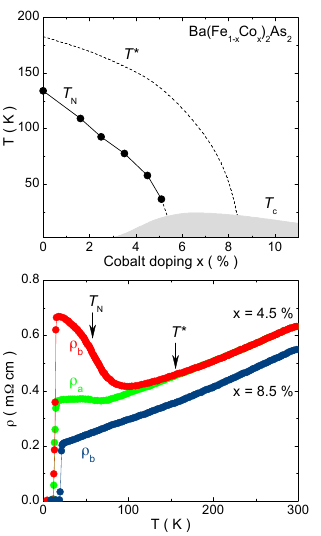}
\caption{Top: Temperature-doping phase diagram of the iron-pnictide superconductor Co-Ba122,
reproduced from Ref.~\cite{ChuScience2010}. 
$T_{\rm N}$ (black dots) marks the onset of SDW (antiferromagnetic) order and $T_c$ the superconducting phase (grey dome). 
$T^{\star}$ marks the approximate onset of the in-plane resistivity anisotropy (see bottom panel). 
Bottom: In-plane electrical resistivity $\rho$ as a function of temperature for Co-Ba122 at $x=4.5$
and 8.5\%~\cite{ChuScience2010}. 
At 4.5\%, there is an upturn in the $b$-axis resistivity $\rho_b$ (red curve), which becomes much larger than that along the $a$-axis ($\rho_a$, green curve). 
This in-plane resistivity anisotropy begins at $T^{\star}$, well above the temperature $T_{\rm N}$ for SDW order. 
At 8.5\%, the electrical resistivity is the same along both axes and is linear in temperature below about 150~K.
}
\label{FigA} 
\end{figure}

A similar precursor effect has been observed in the iron-based superconductor Co-Ba122.
As shown in Figure~\ref{FigA}, the resistivity shows a gradual upturn which begins well above the SDW ordering temperature $T_{\rm N}$,
as a clear deviation from a linear-$T$ dependence at high temperature~\cite{ChuScience2010}.
Moreover, this upturn, a signature of Fermi-surface reconstruction by spin-stripe order, is highly anisotropic in the plane~\cite{ChuScience2010,TanatarPRB2010}, 
reflecting the unidirectional character of the underlying SDW order. The precursor regime is said to have strong nematic character. It is understood theoretically as coming from fluctuations of the unidirectional SDW order~\cite{FernandesPRL2011}.


In YBCO, the temperature below which $\rho(T)$ deviates from its linear-$T$ dependence at high temperature is the standard 
definition of the pseudogap temperature $T^{\star}$~\cite{AndoPRL2004,ItoPRL1993}, so that $T_{\rho} \equiv T^{\star}$.
%
%
%
In Figure~\ref{Fig1}, we plot the temperature $T^{\star}$ extracted from resistivity data on Nd-LSCO and Eu-LSCO as a function of doping, 
which defines the pseudogap phase, using the same criterion. 
We then turn to the Nernst effect. 
In Figure~\ref{Fig3}, we show the Nernst coefficient $\nu=N/H$, where $N$ is the Nernst voltage and $H$ is the magnetic field, 
as a function of temperature in Nd-LSCO and Eu-LSCO for a range of dopings~\cite{OCCNature2009}. 
At high temperature, $\nu/T$ exhibits a weak linear temperature dependence and upon cooling it suddenly rises below a temperature labeled $T_{\nu}$, 
which we trace as a function of doping in the phase diagram of Eu/Nd-LSCO (Figure~\ref{Fig1}). 
Within our error bars, $T_{\nu} = T^{\star}$, so that it is the onset of the pseudogap phase which causes the enhancement of the Nernst coefficient. 
At the QCP for stripe order in Nd-LSCO, at $p^{\star}=0.24$, the Nernst coefficient has a weak and monotonic temperature dependence (Figure~\ref{Fig3}), 
showing no anomaly down to $T\rightarrow0$, the standard behavior of a metal in the absence of FSR.
It is expected theoretically that the FSR caused by stripe order (or SDW order) will enhance the quasiparticle Nernst signal~\cite{HacklPRB2010}.

In YBCO, the Nernst coefficient also displays an enhancement at $T_{\nu}$, 
with $\nu/T$ going from small and positive at high temperature to large and negative at low temperature~\cite{DaouNature2010}. 
(Note that the quasiparticle Nernst signal can be either negative or positive, and this does not directly relate to the sign
of the carriers~\cite{BehniaJPCM2009}.)
In Figure~\ref{Fig1}, we trace $T_{\nu}$ as a function of doping together with $T^{\star}$ extracted from resistivity data, 
showing that $T_{\nu}=T^{\star}$ also holds in YBCO. 
Crucially, the drop in $\nu/T$ at $T^{\star}$ in YBCO is accompanied by a large anisotropy between the $a$ and $b$ axes of the orthorhombic crystal structure, 
whereby $|\nu_a| \ll |\nu_b|$~\cite{DaouNature2010}.
This shows that the pseudogap phase is characterized by a strong nematic tendency, also detected in the resistivity~\cite{AndoPRL2002} and
spin fluctuation spectrum~\cite{HinkovScience2008}. 
By analogy with the iron-pnictide Co-Ba122, where spin-stripe order is preceded by a broad precursor regime of nematicity, 
it may then be that the pseudogap phase in YBCO is a broad precursor of stripe order, as is the case in Nd-LSCO and 
Eu-LSCO~\cite{TailleferJPCM2009,TailleferAR2010}.
It is expected theoretically that a nematic phase will cause a large anisotropy in the quasiparticle Nernst signal~\cite{HacklPRB2009}.
It has also been shown that strong correlations close to a Mott transition can lead to large transport anisotropies in a weakly orthorhombic system such as YBCO~\cite{Okamoto2010}.

\section{Onset of FSR}
\label{Onset}

\begin{figure}[h]
\center		 		
\includegraphics[scale=2.2]{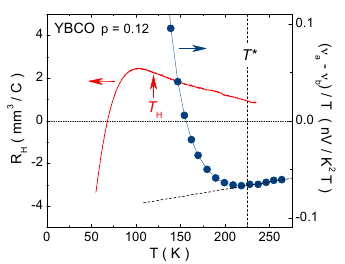}
\caption{Hall coefficient $R_{\rm H}$ versus temperature for YBCO at $p=0.12$, 
measured in a field $H=10$~T (continuous red curve, left axis). 
Below about 100 K, $R_{\rm H}$ drops precipitously to reach large negative values at low temperature~\cite{LeBoeufNature2007,LeBoeufPRB2011}. 
We can define the onset of this drop as $T_{\rm H}$, the temperature below which $R_{\rm H}$ acquires downward curvature~\cite{LeBoeufPRB2011}. 
Also shown is the Nernst anisotropy difference ($\nu_a$ - $\nu_b$ )/$T$ of the same material (blue circles, right axis; from~\cite{DaouNature2010}). 
This quantity starts to rise below a temperature $T_{\nu}$ equal to the pseudogap temperature $T^{\star}$ (vertical dashed line). }
\label{Fig4} 
\end{figure}

As the temperature is reduced further, the Hall effect reveals another temperature scale which we call $T_{\rm H}$. In Figure~\ref{Fig4}, we display the normal-state Hall coefficient $R_{\rm H}$ of YBCO at $p=0.12$. At high temperature, $R_{\rm H}(T)$ has a weak temperature dependence with a positive curvature and, as the temperature is reduced, it drops to negative values. We define the onset of this drop as the inflection point where the curvature goes from positive to negative, as shown by the arrow in Figure~\ref{Fig4}. We associate $T_{\rm H}$ with the temperature at which the high-mobility electron pocket first manifests itself. 
After the onset of strong in-plane anisotropy at $T^{\star}$, 
this is the second step in the process that ultimately leads to FSR at low temperature. 
Applying this analysis of $R_{\rm H}$ to our own data and published data~\cite{SegawaPRB2004} on YBCO, we determined $T_{\rm H}$ as a function of 
doping~\cite{LeBoeufPRB2011}. 
As shown in Figure~\ref{Fig1}, $T_{\rm H}$ rises monotonically with underdoping, exhibiting roughly the same doping dependence as $T^{\star}$, with 
$T^{\star} \simeq 2 T_{\rm H}$. 
We note that $T^{\star}$ and $T_{\rm H}$ have a doping dependence distinct from that of $T_{\mathrm{CO}}$, 
which may come from the fact that the lattice is most effective at stabilizing long-range stripe order at $p=1/8$, for reasons of commensuration.


\section{Summary}
\label{Summary}

Most observers will agree that some of the major outstanding questions in the field of cuprate superconductivity are:\\
\\
\indent 1) What is the mechanism of superconductivity?\\
\indent 2) Why does $T_c$ fall below optimal doping?\\
\indent 3) What is the ``strange metal" phase near optimal doping?\\
\indent 4) What is the pseudogap phase in the underdoped region?\\
 \\
Very similar questions arise in the context of the Bechgaard salts and the iron-pnictide superconductors. In these materials, however, the quantum critical point is well identified and comes from the collapse of SDW order, which provides the central organizing principle for their understanding. In this scenario, superconducting pairing comes from the same interaction that causes SDW order. Below the QCP, both phases compete and $T_c$ falls, leading to the familiar dome-shaped superconducting phase. Near the QCP where magnetic fluctuations are the strongest, the metallic behaviour is anomalous and departs from that of a normal Fermi-liquid. Moreover, the mutual reinforcement of pairing and SDW correlations causes non-Fermi liquid behaviour over an extended range of tuning parameters~\cite{BourbonnaisArxiv2012}, such that the linear-$T$ resistivity persists as long as superconductivity is present in the phase diagram, as indicated by the correlation between the coefficient of linear resistivity and the superconducting $T_c$ seen in these systems~\cite{TailleferAR2010,NDLPRB2009}. 
As for the ``pseudogap phase", in the iron-pnictide Co-Ba122 it is a precursor of the spin-stripe ordered phase at lower temperature~\cite{FernandesPRL2011}.

The same organizing principle provides a coherent picture of electron-doped cuprates, with: 1) a well-characterized SDW phase~\cite{MotoyamaNatPhys2007} and QCP~\cite{DaganPRL2004}; 2) a high-temperature precursor of the SDW order, with strong antiferromagnetic correlations~\cite{MotoyamaNatPhys2007}; 3) a linear-$T$ resistivity~\cite{FournierPRL1998} that correlates with the superconducting $T_c$~\cite{JinNature2011}. It should be mentioned that in the electron-doped cuprates, the correlation length of the fluctuating antiferromagnetic order has been shown to match the thermal de Broglie wavelength, this being a necessary condition for the opening of a pseudogap as seen by photoemission~\cite{KyungPRL2004}.

Turning now to hole-doped cuprates, there is growing body of evidence suggesting that stripe order is a generic phenomenon, not one confined to the 
doped La$_2$CuO$_4$ family of materials. Given this, we propose that a QCP for stripe order is the central organizing principle of hole-doped cuprates. We showed how several properties of the cuprates Eu-LSCO, Nd-LSCO and YBCO are naturally understood within this framework, including the QCP underneath the superconducting $T_c$ dome, the Fermi-surface reconstruction, the universal linear-$T$ resistivity, the correlation between scattering and pairing, and the nematic character of the pseudogap phase. Investigations of other hole-doped cuprates are needed to further establish the universal applicability of a scenario based on a stripe QCP.

While we have focused on the fundamental similarities of organic, pnictide and cuprate superconductors, which we attribute to a common underlying principle, there are also significant differences. It will be of great interest to investigate the role and importance of these differences, including:  the unidirectional/nematic character of the SDW order in pnictides and hole-doped cuprates -- the superconductors with the highest $T_c$; the fact that the dominant modulation in hole-doped cuprates appears to be charge order rather than spin order; the presence of a Mott insulator at low doping in electron-doped and hole-doped cuprates; the presence of an unusual magnetic order in the pseudogap phase of hole-doped cuprates that does not break translational symmetry~\cite{FauquePRL2006,LiNature2008}.

\section*{Acknowledgements}

We wish to thank our collaborators on transport studies of cuprates: 
Luis Balicas, Kamran Behnia, Doug Bonn, Johan Chang, Olivier Cyr-Choini\`ere, Ramzy Daou, Patrick Fournier, Ga\"el Grissonnanche, Walter Hardy, Elena Hassinger, Nigel Hussey, Francis Lalibert\'e, 
David LeBoeuf, Shiyan Li, Ruixing Liang, Liam Malone, Cyril Proust, Brad Ramshaw, Jean-Philippe Reid, Samuel Ren\'e de Cotret, Ilya Sheikin, Michael Sutherland, Hidenori Takagi, Baptiste Vignolle, and Jianshi Zhou.
We acknowledge stimulating discussions with Claude Bourbonnais, Sudip Chakravarty, Andrey Chubukov, Rafael Fernandes, Richard Greene, Marc-Henri Julien, Catherine Kallin, Yong Baek Kim, Hae-Young Kee, Steve Kivelson, Gilbert Lonzarich, Andrew Millis, Michael Norman, Subir Sachdev, Doug Scalapino, Todadri Senthil, Andr\'e-Marie Tremblay, and Matthias Vojta. 
LT acknowledges the long-term support of the Canadian Institute for Advanced Research and funding from NSERC, FQRNT, CFI and a Canada Research Chair.\\





\bibliographystyle{apsrev}
\bibliography{Nematic_ref}







\end{document}